\documentclass{procmult}
\usepackage{graphicx}
\usepackage{amsmath}
\usepackage{wasysym}
\usepackage{units}

\newcommand{\p}{\partial}
\newcommand{\reff}[1]{(\ref{#1})}

\begin{document}

\title{Perspectives in Cosmology, Gravitation and Multidimensions}
\author{Giovanni Montani$^*$, Nakia Carlevaro$^\dag$, Francesco Cianfrani$^\ddag$ \and Valentino Lacquaniti$^\S$}
\institute{
$^*$ ICRA, c/o Dep. of Physics, ``Sapienza'' University of Rome -- Dep. of Physics, ``Sapienza'' Universit\`a di Roma, Piazza A. Moro, 5 (00185), Roma, Italy -- ENEA, C.R. Frascati (Dep. F.P.N.) -- ICRANet, C. C. Pescara [montani@icra.it]\\
$^\dag$  Dep. of Physics, Universit\`a di Firenze -- INFN Florence, Via G. Sansone, 1 (50019), Sesto Fiorentino (FI), Italy -- ICRA, c/o Dep. of Physics, ``Sapienza'' University of Rome [nakia.carlevaro@icra.it]\\
$^\ddag$ ICRA, c/o Dep. of Physics, ``Sapienza'' University of Rome -- Dep. of Physics, ``Sapienza'' Universit\`a di Roma, Piazza A. Moro, 5 (00185), Roma, Italy [cianfrani@icra.it] \\
$^\S$ Dep. of Physics, ``E. Amaldi'', University of Rome ``Roma Tre'', Via della Vasca Navale 84, I-00146, Rome, Italy -- ICRA, c/o Dep. of Physics, ``Sapienza'' University of Rome -- LAPTH-9, Chemin de Bellevue BP, 110, 74941 Annecy Le Vieux Cedex, France [lacquaniti@fis.uniroma3.it]
}
\maketitle
\abstract
Recent developments from the activity of the CGM Group are discussed. Cosmological implications of fundamental approaches to quantization of gravity are presented in order to fix the main issues as well as perspectives for future investigations. Particular attention will be devoted to the classical and quantum features of the generic inhomogeneous Universe, to the role of reference frame in quantum gravity, and eventually to phenomenological features related with the Kaluza-Klein framework. 

\section{Introduction}
Since September 2005, we gave to our research
activity the form of a compact group working on
topics related to specific features in Cosmology,
Gravitation and multidimensions (named the CGM Group).
This group is developing its investigation efforts
within the II Chair in Theoretical Physics
(Physics Department of University of Rome ``Sapienza''),
devoted to the study of Relativistic Astrophysics.
The CGM Group, directed by Prof. Giovanni Montani,
was constituted by over these years by about ten
Students and Postdocs, having different educational
levels and research interests, but sharing the
common aim to jointly pursue the study of subtle
questions concerning the Universe birth and evolution.

This work presents a limited sample of the theoretical
analyses we developed in the three fields we mentioned above.
In particular, we address here the idea of an evolutionary
quantum gravity, the efforts aimed to remove the use
of the ``time gauge'' in Loop Quantum Gravity,
the evolution of the early Universe in presence
of bulk viscosity effects and eventually, the proposal
for a phenomenological approach, solving some
shortcomings of the Kaluza-Klein paradigm.

In order to call attention on the
experience related to the CGM Group project, we list here
the names, in alphabetic order, of the participants in it involved
over these years:
Dr Marco Valerio Battisti,
Dott. Riccardo Belvedere, Dr Riccardo Benini,
Dr Nakia Carlevaro, Dott. Michele Castellana,
Dr Francesco Cianfrani, Dr Giovanni Imponente,
Dr Valentino Lacquaniti, Dr Orchidea Maria Lecian,
Dr Simone Mercuri, Dott. Irene Milillo,
Dott. Francesco Vietri, Dott. Simone Zonetti.

\section{Gravity in a synchronous reference frame}\label{1o}
We are going to analyze the implications of restricting the admissible set of coordinate transformations to those ones preserving the space-time splitting of the full 4-dimensional manifold.

In this respect, let us consider a $3+1$ slicing of space-time coordinates $y^\mu$, {\it i.e.} $y^\mu=y^\mu(t,x^i)$, $t$ and $x^i$ ($i=1,2,3$) being the time-like and the space-like coordinates, respectively. The above mentioned restriction can be performed by fixing a synchronous (gaussian) reference frame, {\it i.e.} 
\begin{equation}
g_{00} = 1,\qquad g_{0i} = 0.
\end{equation}

The transformations preserving the conditions above are the following ones
\begin{eqnarray}
\label{srp3}
t^{\prime } = t + \xi (x^l)\\
x^{i^{\prime }} = x^i+\xi^i= x^i
+ \partial _j\xi \int h^{ij}dt + \phi ^i(x^l)
\, , \label{srp3bis}
\end{eqnarray}

where $\xi=\xi(x^j)$ and $\phi^i=\phi^i(x^j)$ denote generic space functions.
 
The Lagrangian of the gravitational field, in presence of a cosmological constant $\Lambda$, can be written as
\begin{equation}
L_{grav} = \int _{\Sigma ^3_t}d^3x \mathcal{L}_{grav} = 
-\frac{1}{2c^2k}\int _{\Sigma ^3_t}
d^3x\sqrt{h} \left\{ K^2 - K_{ij}K^{ij} - ~^3R\right\} -\frac{1}{k}\int _{\Sigma ^3_t}
d^3x\sqrt{h}\Lambda 
\, ,
\label{srp5}
\end{equation}

where $k$ is the Einstein constant
($k = 8\pi G/c^4$), $h\equiv det h_{ij}$ and
$K_{ij}$
refers to the extrinsic curvature.

The invariance of the Lagrangian density under the set of transformations (\ref{srp3bis}) implies the following conditions
\begin{eqnarray}
\delta L =
\frac{\delta L}{\delta h_{ij}}\delta h_{ij} + 
\frac{\delta L}{\delta (\partial _l h_{ij})}
\delta (\partial _l h_{ij}) + \frac{\delta L}{\delta (\partial _t h_{ij})}
\delta (\partial _t h_{ij}) + \int \partial _t \mathcal{L}\xi d^3x = 0 
\, ,
\label{srp7i}
\end{eqnarray}

from which, by using the Euler-Lagrange equations for the proposed system, the conservation law below is obtained \cite{EQ}

\begin{equation}
\partial _t\left\{ \int _{\Sigma ^3_t}d^3x\left[ 
\frac{\delta \mathcal{L}}{\delta (\partial _th_{ij})}
\delta h_{ij} + \mathcal{L}\xi \right] \right\} = 0 
\, . 
\label{srp11}
\end{equation} 

By an integration by parts this expression can be written as
\begin{equation}
\int _{\Sigma ^3_t}d^3x\left\{ 
-\partial _t(H_i)\left( \phi^i + \partial _l\xi \int dt h^{il}
\right) - \left( \partial _tH - \partial _lH^l\right) \xi \right\} = 0
\, ,
\label{srp14}
\end{equation}

and it outlines the emergence of the following constraints
\begin{equation}
\partial _tH_i = 0 \; , \; \partial _tH = \partial _lH^l 
\, .
\label{srp15}
\end{equation}

The first three conditions imply $H_i = k_i(x^l)$, such that the last one gives $\partial _tH = \partial _lk^l$. Since the invariance 
under 3-diffeomorphisms holds, we expect that the super-momentum constraints are not modified. Hence we find 
\begin{equation}
H^* \equiv H -\mathcal{E}(x^i)= 0 \, \quad H_i = 0 
\, .
\label{srp16}
\end{equation}

Therefore, in order to preserve the synchronous character of the reference, the super-Hamiltonian must be non-vanishing. 

Within this scheme we can write $\mathcal{E}(x^i)=-2\sqrt{h}\rho$ and the function $\rho$ can be seen as the energy density associated to a co-moving dust fluid. Indeed such an interpretation implies the requirement that the energy density $\rho$ is positive. As soon as this condition holds, this additional contribution to Einstein equations provides the physical realization of the synchronous gauge. This way, a clear correspondence between fixing the frame and introducing a matter field is established.

The function $\rho$ can be investigated in a Wheeler-DeWitt framework, where $H$ and $H_i$ are promoted to operators and conditions (\ref{srp16}) are implemented by the Dirac prescription on physical states.

In particular, this formulation gives naturally an evolutionary paradigm for the quantum gravitational field, in fact the associated Schr\"odinger equation reads as follows
\begin{equation}
i\hbar \partial _t\psi =
\int _{\Sigma ^3_t}\hat{H} d^3x  \psi= \int _{\Sigma ^3_t}\mathcal{E} d^3x  \psi
\, .
\end{equation}

Therefore, the wave-functionals evolve with the label time $t$ and this solves the problem of time in Quantum Gravity. Furthermore, $\mathcal{E}(x^i)$ is fixed by the initial conditions we assign on a non-singular hypersurfaces. 

The implementation of this scheme in an inhomogeneous Universe demonstrates that this scenario works properly in a cosmological setting \cite{EQ}. In fact, there exists a negative portion of the $H$ spectrum, which turns out to be bounded from below as soon as a discrete spatial structure is imposed. Hence, if we assume the Universe to be in the fundamental state, $\rho$ is positive and an estimate for it can be given. By fixing the minimum length as the Planck scale, we obtain the following estimate for $\Omega_\rho$
\begin{equation}
\Omega _\rho \equiv
\mathcal{O}\left( 10^{-60}\right)
\, .
\label{crit}
\end{equation}  

This estimate outlines that this dust fluid contribution has a negligible effect on the classical Universe dynamics. Hence, this scenario predicts no modification with respect to the standard cosmological scenario, while the quantum dynamics is significantly affected. This result is consistent with restoring General Relativity in the classical limit.

The cosmological analysis can be enriched by introducing a scalar field playing the role of the inflaton. The set of constraints (\ref{srp16}) is simply modified by the addition of the super-Hamiltonian and of the super-momentum of the scalar field. 

Furthermore, a proper quantum to classical transition can be properly addressed. However, as in standard quantum cosmology, there are still problems in describing the complete evolution of the Universe and, in particular, in dealing with spatial gradients of anisotropy parameters. 

An analogous conclusion concerning the appearance of a non-vanishing super-Hamiltonian eigen-value holds for a stochastic gravitational field as soon as an ensemble description is addressed from a fundamental quantum picture \cite{EQ}. In particular, if it is fixed a fundamental correspondence between the ensemble dynamics of stochastic gravitational systems and the semi-classical WKB limit of their quantum dynamics, it can be shown that the non-stationary character of the ensemble distribution reflects the existence of a non-zero super-Hamiltonian eigenvalue of order $\hbar$. The application of this framework to the inhomogeneous Universe outlines that a Schr\"odinger-like dynamics is predicted when a portion of
the system de-parametrizes from the whole and it plays the role of a good time variable.  

\section{Local Lorentz invariance in canonical Quantum Gravity}

Let us investigate the local Lorentz invariance in a second-order formulation, which means that combinations of 4-bein components are taken as configuration variables. In particular, generic basis of the tangent space can be written as
\begin{equation}
e^0_\mu=(N,\chi_a E^a_i),\qquad e^a_\mu=(E^a_iN^i, E^a_i).\label{tetr}
\end{equation} 

It is worth noting the role of $\chi_a$, which give the velocity components of the 4-bein frame with respect to one co-moving with equal-time hypersurfaces. The lapse function $\tilde{N}$ and the shift vector $\tilde{N}^i$ can be expressed in terms of $N$, $N^i$, $\chi_a$ and $E^a_i$ \cite{FG07}.

We perform the Hamiltonian formulation of the Einstein-Hilbert action in the phase space parametrized by configuration variables $\tilde{N}, \tilde{N}^i, E^a_i, \chi_a$ and associated conjugate momenta ($\pi_{\tilde{N}}$, $\pi_i$, $\pi^i_a$ and $\pi^a$, respectively). The total Hamiltonian is a linear combination of constraints, {\it i.e.}
\begin{equation}
\mathcal{H}=\tilde{N}'H+\tilde{N}^iH_i+\lambda^{\tilde{N}}\pi_{\tilde{N}}+\lambda^i\pi_i+\eta_a\varphi^a+\lambda_a\Phi^a,
\end{equation}

where $ \lambda^{\tilde{N}},\lambda^i,\lambda_a,\eta_a$, $ \tilde{N}'=\sqrt{h}\tilde{N}$ and $N^i$ are Lagrangian multipliers. While the vanishing of the super-Hamiltonian $H$ and the super-momentum $H_i$ gives the invariance under spatial diffeomorphisms and time re-parametrizations, respectively, 
the local Lorentz invariance is insured by the conditions $\Phi^a=0$ and $\varphi^a=0$, {\it i.e.}  
\begin{eqnarray}
\Phi^a=T^{-1a}_{\phantom1b}(\chi)\pi^b+\delta^{ab}\pi^i_b\chi_cE^c_i=0,\qquad
\varphi^a=\epsilon_{\phantom1b}^{a\phantom1c}(\pi^b\chi_{c}-\pi^i_{c}E^b_i)=0,
\end{eqnarray}

$T^{-1a}_{\phantom1b}$ being $\delta^a_b+\chi^a\chi_b$.

The investigation of the local boost invariance is performed by fixing an arbitrary local Lorentz frame and searching for a unitary operator mapping the Hilbert spaces corresponding to different frames.

In particular, we fix $\chi_a$ equal to some space-time functions $\bar{\chi}_a(t,x^i)$ and we eliminate $\pi^a$ by solving the boost constraints  $\Phi^a=0$. This allow us to write the following action in the adopted frame reads 
\begin{equation}
  S=-\frac{1}{16\pi G}\int[\pi^i_a\partial_tE^a_i+\pi_{\tilde{N}'}\partial_t\tilde{N}'+\pi_i\partial_t\tilde{N}^i-\tilde{N}'H^{\bar\chi}-\tilde{N}^iH^{\bar\chi}_i-\lambda_a\varphi_{\bar{\chi}}^a-\lambda^{\tilde{N}}\pi_{\tilde{N}}-\lambda^i\pi_i]dtd^3x,
\end{equation}

where $H^{\bar{\chi}}=H(\chi_a=\bar{\chi}_a)$ and $H^{\bar{\chi}}_i=H_i(\chi_a=\bar{\chi}_a)$, while now the rotation constraints read
\begin{eqnarray}
 \varphi_{\bar{\chi}}^a=\epsilon^{abc}(\bar{\chi}_b\pi^i_dE^d_i\bar{\chi}_d-\delta_{db}\pi^i_{c}E^d_i).
\end{eqnarray}

A formal quantization can be defined by promoting $E^a_i$, $\tilde{N}$ and $\tilde{N}^i$ to multiplicative operators and replacing Poisson brackets with commutators. Indeed a proper Hilbert space cannot be defined (we are working in a Wheeler-DeWitt framework), nevertheless we expect the discussion below to be implementable in a rigorous quantum framework once it would be available.  
 
Let us now restrict to the sector $\bar{\chi}_a=0$. The boost operator in the $(E^a_i,\pi^j_b)$ phase space can be represented by the unitary operator
\begin{equation}
  U_\epsilon=I-\frac{i}{4}\int\epsilon^a\epsilon_b(E^b_i\pi^i_a+\pi^i_aE^b_i)d^3x+O(\epsilon^4),
\end{equation}

which map 3-bein vectors between the two sectors $\bar{\chi}_a=0$ and $\bar{\chi}_a=\epsilon_a$.

In order to realize a quantum symmetry, the unitary operator $U$ must send physical states into physical states. As for the super-Hamiltonian and the super-momentum, this property is obvious since we have at the $\epsilon^2$ order that
\begin{equation}
U_\epsilon H^0U_\epsilon^{-1}=H^{\epsilon}\hspace{0.4cm}U_\epsilon H^0_iU_\epsilon^{-1}=H^{\epsilon}_i.
\end{equation}

As soon as the rotation constraints are concerned, one finds that it is not mapped into the corresponding expression for $\bar{\chi}_a=\epsilon$. However, with some algebra it can be shown \cite{FG07} that the transformed state is annihilated by $\varphi_\epsilon^a$.

Therefore, $U$ maps physical states between the two sectors $\bar{\chi}=0$ and $\bar{\chi}=\epsilon$. Hence, although the Lorentz frame has been fixed before the quantization, nevertheless we expect to be able to implement the boost symmetry within a Wheeler-DeWitt framework in a second order formulation.

Then, we consider a first-order formulation, where 4-bein components and spin connections are treated as independent variables. In this case we have at our disposal a well-defined quantization procedure at least at the kinematical level: Loop Quantum Gravity \cite{revloop}. Such a procedure is based on using techniques proper of lattice gauge theories and, in this respect, a reformulation of gravity is performed within the time gauge by which some SU(2) connections (Ashtekar-Barbero-Immirzi variables) come out, whose canonically conjugate variables are densitized 3-bein vectors of the spatial metric. We are going to outline that such a restriction to the time-gauge case is not necessary and that the boost invariance can be implemented on a quantum level. 

Let us start with the Holst action \cite{Ho96} for gravity, which in units $8\pi G=c=1$ reads as follows
\begin{equation} 
S=\int \sqrt{-g}e^\mu_A e^\nu_B\left( R_{\mu\nu}^{AB}-\frac{1}{2\gamma}\epsilon^{AB}_{\phantom{12}CD}R^{CD}_{\mu\nu}\right)d^4x,\label{act}
\end{equation}

$\omega^{AB}_\mu$ being  spin connections, while
$R^{AB}_{\mu\nu}=\partial_{[\mu}\omega^{AB}_{\nu]}+\omega^A_{\phantom1C[\mu}\omega^{CB}_{\nu]}$. $\gamma$ is the Immirzi parameter. By a Legendre transformation the following Hamiltonian density comes out
\begin{eqnarray}   \mathcal{H}=\int\bigg[\frac{1}{eg^{tt}}H-\frac{g^{ti}}{g^{tt}}H_i-\omega^{AB}_t{}^\gamma\!p^{CD}_{\phantom1\phantom2AB}G_{CD}+\lambda_{ij}C^{ij}+\eta_{ij}D^{ij}+\lambda^{AB}\pi_{AB}^t\bigg]d^3x,\end{eqnarray}

where $1/eg^{tt}$, $g^{ti}/g^{tt}$, ${}^\gamma\!p^{CD}_{\phantom1\phantom2AB}\omega^{AB}_t$, $\lambda_{ij}$, $\eta_{ij}$ and $\lambda^{AB}$ are Lagrangian multipliers. Here $H$ and $H_i$ denote the super-Hamiltonian and the super-momentum, respectively, while $G_{AB}$ are Gauss constraints of the Lorentz symmetry. $C^{ij}$ and $D^{ij}$ are not associated with any gauge symmetry and in fact they made the whole system of constraints second-class.
We provided \cite{prl} a solution to $C^{ij}=D^{lk}=0$ such that the reduced phase space can be parametrized by some variables $\{\widetilde{A}^a_i,\chi_b\}$ and their canonically conjugate variables $\{\widetilde{\pi}_a^i,\pi^b\}$, respectively. Functions $\chi_a$ have the same geometrical interpretations as in the second-order case, while, since $\widetilde{\pi}_a^i$ are densitized inverses 3-bein vectors of the spatial metric, $\widetilde{A}^a_i$ are the extension of Ashtekar-Barbero-Immirzi to a generic Lorentz frame. The full action reads as 
\begin{equation}
S=\int d^4x\bigg[\widetilde{\pi}^i_a\partial_t\widetilde{A}^a_i+\pi^a\partial_t\chi_a-\frac{1}{\sqrt{g}g^{tt}}H+\frac{g^{ti}}{g^{tt}}H_i+\eta^aG_a+\lambda^aB_a\bigg],\label{FINALACT}
\end{equation}

$\lambda^a$ and $\eta^a$ being Lagrangian multipliers, while $G_a$ are Gauss constraints of the SU(2) group, whose connections are $\widetilde{A}^a_i$ themselves. Therefore, the phase space of General Relativity resembles that of an SU(2) gauge theory also when the time-gauge condition does not hold and the quantization techniques proper of LQG can be applied.

As for $B^a$, they simply give the vanishing of $\pi^a$ and they can be implemented on a quantum level taking wave-functionals not depending on $\chi_a$ variables. 

Within this scheme, boost transformations are generated by a linear combination of $G_a$ and $B_a$, thus physical states must be invariant under boosts. 

\section{The Quasi-Isotropic Model}
In 1963, E.M. Lifshitz and I.M. Khalatnikov \cite{lk63} proposed the so-called Quasi-Isotropic (QI) Solution. This model is based on the idea that a Taylor expansion, in time, of the 3-metric can be addressed. In this approach, the pure isotropic solution becomes a particular case of a larger class of solutions, existing only for space filled with matter. In fact, for an ultra-relativistic perfect fluid, \emph{i.e.}, $p=\rho/3$ ($\rho$ being the fluid energy-density and $p$ the thermostatic pressure, respectively), the spatial metric assumes the form $h_{ij}\sim a_{ij}(x^k)\,t$. 
As a function of time, the 3-metric is expandable in powers of $t$. The QI Solution is formulated in a synchronous system, with a line element of the following form
\begin{equation}
ds^2 = - dt^2 + h_{ij}(t, x^k)dx^{i}dx^{j}\;,\qquad
h_{ij}=t\;{a}_{ij}(x^k)+t^{2}\;{b}_{ij}(x^k)+...\;.
\end{equation}

To include dissipative effects into the evolution of the energy source \cite{nkmIJMPD}, we deal with a more complex (no integer powers) 3-metric expansion:
\begin{equation}
h_{ij}=t^{x}\;{a}_{ij}+t^{y}\;{b}_{ij}+...\;,\;\qquad\quad
h^{ij}=t^{-x}\;{a}^{ij}-t^{y-2x}\;{b}^{ij}+...\;,
\end{equation}
where we impose the constraint guaranteeing the space contraction, $\;x>0$, and the consistence of the perturbative scheme, $y>x$. We propose here the immediate generalization of the original QI scheme by considering the presence of dissipative processes within the asymptotic dynamics, as expected at temperatures above $\mathcal{O}(10^{16} GeV)$. This extension is described by an additional term in the expression of the perfect-fluid energy-momentum tensor:
\begin{equation}
T_{\mu\nu}= \tfrac{1}{3}\;\rho\;(4u_{\mu}u_{\nu}+g_{\mu\nu})-
\zeta\,u^{\rho}_{\,;\,\rho}(u_{\mu}u_{\nu}+g_{\mu\nu})\;,\qquad
u^{\rho}_{\,;\,\rho}=\p_t\ln\sqrt{h}\;,
\end{equation}
where $p=\nicefrac{\rho}{3}$ is the thermostatic pressure, $u_\mu$ is the fluid 4-velocity and $\zeta$ is the \emph{bulk viscosity} coefficient. In particular, here we assume this quantity as a function of the Universe energy-density $\rho$; according to literature developments, we express $\zeta$ as a power-law of the form $\zeta=\zeta_0\,\rho^s$, where $\zeta_0=const.$ and $s$ denotes a dimensionless parameter ($0\leq s\leq\nicefrac{1}{2}$) \cite{bk77}. In what follows, we fix the value $s=\nicefrac{1}{2}$ in order to deal with the maximum effect that bulk viscosity can have without dominating the dynamics \cite{barrow88}, in view of the phenomenological issue of thermal equilibrium perturbations which characterizes this kind of viscosity.

Writing now the \emph{00}-component of Einstein Eqs., we can expand the energy density as
\begin{equation}\label{pippo}
\rho=\frac{e_0}{t^2}\;+\;\frac{e_1\,b}{t^{2-y+x}}\;,\;\qquad
\sqrt{\rho}=\frac{\sqrt{e_0}}{t}\left(1\;+\;\frac{e_1\,b}{2e_0}\,t^{y-x}\right)\;,
\end{equation}
where we have assumed the condition $\;u_0^2\simeq1$\; (with $u_0=-1$) whose consistence must be verified \emph{a posteriori} and $e_0$, $e_1=const.$

\textbf{The Viscous Solutions} - To obtain an analytical solution, we match the \emph{00}-Einstein Eq., after substituting eq. \reff{pippo}, with the hydrodynamical ones $T_{\mu;\,\nu}^{\nu}=0$. Under our hypothesis ($u_\alpha$ is neglected wrt $u_0$), these Eqs. can be combined together and solved order by order in the $1/t$ expansion (asymptotically as $t\rightarrow0$), obtaining the following solutions:
\begin{equation}
\label{0-sol}
x=1/[1-\tfrac{3\sqrt{3}}{4}\,\zeta_0]\;,\quad
e_0=\tfrac{3}{4}\;x^{2}\;,\qquad\quad
y=2\;,\quad
e_1=-\tfrac{1}{2}\,x^{3}+2x^{2}-2x\;.
\end{equation}

The parameter $\zeta_0$ has the restriction $\zeta_0\leq\nicefrac{4}{3\sqrt{3}}$ in order to satisfy $x>0$, this way the exponent $x$ runs from $1$ (which corresponds to $\zeta_0=0$) to $\infty$. We remark that this constraint arises from a zeroth-order analysis and defines the existence of a viscous Friedmann-like model. We now narrow the validity of $x$ to the values which satisfy the constraint $x<y$, guaranteeing the consistence of the model. From \reff{0-sol}, the QI Solution exists only if
\begin{equation}
\label{zzero}
\zeta_0<\zeta_0^{*}=2/3\sqrt{3}\;,
\end{equation}
\emph{i.e.}, the viscosity is sufficiently small. For values of the viscous parameter $\zeta_0$ that overcome the critical one $\zeta_0^{*}$, the asymptotic QI Solution can not be addressed, since perturbations would grow more rapidly than zeroth-order terms. We are now able to write down the final expression for $\rho$ and for the density contrast $\delta$, defined as the first- and zeroth-order ratio:
\begin{equation}
\rho=\frac{3\,x^{2}}{4\,t2}\;-\;\frac{(x^{3}/2-2x^{2}+2x)\;b}{t^{x}}\;,\qquad
\delta\,=\,-\tfrac{8}{3}\,(\tfrac{1}{4}\;x+\tfrac{1}{x}-1)\,b\;t^{2-x}\;,\qquad
1\leq x<2\;.
\end{equation}
We note that the density-contrast evolution is strongly damped by the presence of dissipative effects: this behavior implies that $\delta$ approaches the singularity more weakly as $t\rightarrow0$ when the viscosity coefficients runs to $\zeta_0^{*}$. In correspondence to this threshold value, $\delta$ remains constant in time and hence it must be excluded by the possible $\zeta_0$ choices.    We conclude this Section verifying the consistence of our model. By the analysis of $\alpha$\emph{0}-gravitational Eq. we get, up to the leading-order of expansion: $u_\alpha\sim\;t^{3-x}$. The assumption $u_0^{2}\simeq1$ is, therefore, well verified. In fact, $u_\alpha u^\beta\sim t^{6-3x}$ can be neglected in the contraction $u_\mu u^\mu=-1$ and our approximation scheme results to be self-consistent.

\textbf{Comments on the adopted paradigm} - It is well known \cite{kolb} the crucial role played in cosmology by the \emph{microphysical horizon}, as far as the thermodynamical equilibrium is concerned. In the isotropic Universe, such a quantity is fixed by the inverse of the expansion rate, $H^{-1}\equiv (R/\dot{R})$ ($R$ being the scale factor of the Universe and the dot identifies time derivatives) and it gives the characteristic scale below which the elementary-particle interactions are able to preserve the thermal equilibrium of the system. Therefore, if the mean free-path of particles $\ell$ is greater than the microphysical horizon (\emph{i.e.}, $\ell>H^{-1}$), no real notion of thermal equilibrium can be recovered at the micro-causal scale. If we indicate by $n$ the number density of particles and by $\sigma$ the averaged cross section of interactions, then the mean free-path of the ultra-relativistic cosmological fluid (in the early Universe the particles velocity is very close to speed of light) takes the form
$\ell\sim 1/n\sigma$. Interactions mediated by massless gauge bosons are characterized by the cross section $\sigma\sim\alpha^{2}\,T^{-2}$ ($g=\sqrt{4\pi\alpha}$ being the gauge coupling strength) and the physical estimation $n\sim T^{3}$ leads to the result $\ell\sim1/\alpha^{2}T$ \cite{kolb}. During the radiation-dominated era $H\sim T^{2}/m_{Pl}$, so that
\begin{equation}
\ell\sim \frac{T}{\alpha^2m_{Pl}}\;H^{-1}\;.
\end{equation}

Therefore in the case of $T\apprge\alpha^2m_{Pl}\sim \mathcal{O}(10^{16}GeV)$, \emph{i.e.}, during the earliest epochs of pre-inflating Universe, the interactions above are effectively ``frozen out'' and they are not able to maintain or to establish thermal equilibrium. As a consequence of this non-equilibrium configuration of the causal regions characterizing the early Universe, most of the well-established results about the kinetic theory \cite{weinberg71} concerning the cosmological fluid nearby equilibrium become not applicable. Indeed all these analysis are based on the assumption to deal with a finite mean free-path of the particles and, in particular, results about the characterization of viscosity are established when pure collisions among particles are retained. However, when the mean free-path is grater than the micro-causal horizon (which, in the pre-inflating Universe, coincides with the cosmological horizon), $\ell$ can be taken of infinite magnitude for any physical purpose. This way, our model was essentially based on an \emph{hydrodynamical description}, \emph{i.e.}, assuming that an arbitrary state is adequately specified by the particle-flow vector and the energy momentum tensor alone. Following this point of view, viscosity effects are treated on the ground of a thermodynamical description of the fluid, \emph{i.e.}, the viscosity coefficient is fixed by the macroscopic parameters governing the system evolution.

\section{Matter reduction in Kaluza-Klein model}
It is well known that in the 5D  KK model the  geodesic approach provides unphysical results: it yields an upper bounded $q/m$ ratio for test particles and a tower of  massive modes beyond the Planck scale \cite{overduinwesson}. Here we propose a new approach, where we start by Einstein 5D equation in presence of 5D matter: ${}^5R^{AB}=8\pi G_{5}T^{AB}$. Via the dimensional reduction we get the equations \cite{KK2}

\begin{eqnarray}
\label{1}
&{}& G^{\mu\nu}=\frac{1}{\phi}\nabla^{\mu}\partial^{\nu}\phi-\frac{1}{\phi}g^{\mu\nu}g^{\alpha\beta}\nabla_{\alpha}\partial_{\beta}\phi+8\pi G\phi^2T^{\mu\nu}_{em}+8\pi G \frac{T^{\mu\nu}_{matter}}{\phi} ,\\
\label{2}
 &{}& 
  \nabla_{\nu}\left(\phi^3F^{\nu\mu}\right)=4\pi j^{\mu}, \\
  \label{3}
 &{}& 
g^{\alpha\beta}\nabla_{\alpha}\partial_{\beta}\phi=-G \phi^3 F^{\mu\nu}F_{\mu\nu}+\frac83 \pi G \left( T_{matter}+2\frac{\vartheta}{\phi^2}\right),
  \end{eqnarray} where, being $l_5$ the coordinate length of the extra dimension, we define $
  T^{\mu\nu}_{matter}=l_5\phi T^{\mu\nu} , j^{\mu}=\sqrt{4G}l_5\phi T_5^{\mu}, \vartheta=l_5\phi T_{55} , G=G_5l_5^{-1}$.
  We also get, by mean of Bianchi identities, the equations:
  \begin{equation}
  \nabla_{\rho}( T^{\mu\rho}_{matter})=-g^{\mu\rho}\left(\frac{\partial_{\rho}\phi}{\phi^3}\right)\vartheta+ F^{\mu}_{\,\,\rho}j^{\rho},\quad\quad \nabla_{\mu}j^{\mu}=0.\label{4}
 \end{equation}
 Starting from these equations, and applying a multipole expansions we can define a test particles and analyze its motion \cite{KK1}. The resulting equation is:
$$
m\frac{Du^{\mu}}{Ds}=A(u^{\rho}u^{\mu}-g^{\mu\rho})\frac{\partial_{\rho}\phi}{\phi^3}+qF^{\mu\rho}u_{\rho}.
$$
Coupling factor and the effective tensor for test particles read as follows :
\begin{eqnarray*}
&& m\!=\!\frac{1}{u^0}\int\!\!\!d^3x\,\sqrt{g}\, T^{00}_{matter},\qquad
q\!=\!\int\!\!\!d^3x\,\sqrt{g}\,j^0 ,
\qquad A\!=\!u^0\int\!\!\!d^3x\,\sqrt{g}\, \vartheta, \quad \\
&& \sqrt{g} T^{\mu\nu}_{matter}\!=\!\int\!\!\!ds\,m\,\delta^4(x-X)u^{\mu}u^{\nu},
\qquad  \sqrt{g} j^{\mu}\!=\!\int\!\!\!ds\,q\,\delta^4(x-X)u^{\mu},
\qquad \sqrt{g}\vartheta\!=\!\int\!\!\!ds\,A\,\delta^4(x-X)
\end{eqnarray*}
Here the test particle is not localized along the extra dimension due to the compactification hypothesis while, actually, in the geodesic approach is taken for granted that is possible to deal with a 5D localized particle. Such a  motion equation provides a new scenario with respect to the usual KK one. 
Indeed, studying the effective Action, and the associate Klein-Gordon equation, it is possible to show that such a dynamics does not provide a tower of massive modes. The removal of the KK tower is the most striking result of this approach \cite{KK1}. Therefore, looking at (\ref{1},\ref{2},\ref{3},\ref{4}) we now have a consistent compactified KK model with matter which appears as a theory of gravity modified by the presence of the scalar fields $\phi$ and $\vartheta$, which can be addressed to dark energy and dark matter respectively \cite{KK2}. The possibility to deal with such issues is interesting and first studies of the model in homogeneous and spherical backgrounds - which are in progress-  seem to be encouraging.

\end{document}